Short Communication

# Single molecule study of DNA collision with elliptical nanoposts conveyed by hydrodynamics

Yannick VIERO[1,2], Qihao HE[1,2], Marc Fouet[1,2], and Aurélien BANCAUD[1,2,]*

CNRS, LAAS

* E-mail: abancaud@laas.fr

Dr. Y. Viero, Dr. Q. He, M. Fouet, H. Ranchon, Dr. A. Bancaud

[1] CNRS, LAAS, 7 avenue du colonel Roche, F-31400 Toulouse, France

[2] Univ de Toulouse, LAAS, F-31400 Toulouse, France

Word count: 2570

**Abstract**

Periodic arrays of micro- or nano-pillars constitute solid state matrices with excellent properties for DNA size separation. Nanofabrication technologies offer many solutions to tailor the geometry of obstacle arrays, yet most studies have been conducted with cylinders arranged in hexagonal lattices. In this report, we investigate the dynamics of single DNA collision with elliptical nanoposts using hydrodynamic actuation. Our data shows that the asymmetry of the obstacles has minor effect on unhooking dynamics, and thus confirms recent predictions obtained by Brownian dynamics simulations. In addition, we show that the disengagement dynamics are correctly predicted by models of electrophoresis, and propose that this consistency is associated to the confinement in slit-like channels. We finally conclude that elliptical posts are expected to marginally improve the performances of separation devices.

Size separation of DNA molecules with electrophoresis is a key process of molecular biology, which has been widely investigated over decades [1]. Separation cannot be performed in free solution [2], requiring the use of matrices to achieve size-dependent DNA mobilities. Slab gels are the most common separation matrices, but optimal performances have been reached with capillaries of hundreds of µm in diameter filled with concentrated polymer solutions. This technology features excellent physical properties associated to rapid heat dissipation and reduction of convection [3], and the chemical composition of polymer solutions has been tailored to enhance the resolution of separation experiments, which now reach base-pair (bp) resolution for fragments up to ~500-1000 bp [4]. Despite these successes, the nanoscale arrangement of polymer solutions is intrinsically disordered, making it difficult to precisely control the migration of DNA at the molecular level. Therefore Volkmuth and Austin proposed to use micro or nanofabricated solid state artificial matrices for DNA separation [5]. These systems consist of arrays of posts etched in glass or silicon with exquisite control over their size, spacing, and shape from the nano to the millimeter scale. This approach showed impressive results for long DNA chains of ~$100.10^3$ base pairs (100 kbp) [6] that were separated in tens of seconds in comparison to hours otherwise [7]. For shorter DNA fragments, artificial matrices of nanoposts of ~200 nm in radius have been devised, also showing exquisite performances [8].

Machining technologies offer many choices for the design of the matrix, yet hexagonal arrays of cylindrical obstacles have been predominantly characterized in the literature [9]. The idea of using asymmetric obstacles to separate biomolecules has nevertheless been documented in the context of e.g. Brownian ratchet devices [10], which enable to perform spatial separations of DNA based on the size dependence of its diffusion coefficient [11]. Dense and asymmetric obstacles have also been incorporated into artificial matrices to test their separation power [12], and it was concluded that DNA migration velocity was dependent on the lattice geometry. Because the contribution of obstacle geometry on hooking events was not investigated in the latter study, Brownian dynamics simulations of single DNA molecule colliding with elliptical posts were conducted [13]. This research indicated that the holdup time was relatively insensitive to the orientation of the ellipse, and it was suggested that elliptical obstacles should be less effective than cylindrical obstacles for DNA separations.

In this report, we set out to check this prediction using single molecule microscopy, focusing on DNA collision with small elliptical obstacles of different geometries. We use hydrodynamics to convey DNA molecules in order to disregard the adverse effects of uncontrolled electro-osmotic flows [12]. This actuation principle is also more favorable to monitor DNA conformation during collisions, because hooks, usually referred to as U/J collisions, are observed at an occurrence larger than 90% with hydrodynamics [14], whereas they are less frequent using electrophoresis [15,16]. Therefore we focus on U/J collisions in this report.

Nanoposts were generated using dry etching of silicon after nanopatterning a photoresist by PDMS-based phase-shift lithography (see [17] for details of the fabrication process). The size of elliptical obstacles was set to $D=575+/-75$ nm and $d=190+/-45$ nm in major and minor diameters (Fig. 1A). The height of these obstacles was $0.8+/-0.1$ µm. We chose to use the smallest obstacles we could fabricate regarding the height of the obstacles. This choice was made based on our previous conclusions [14] that showed that the unhooking characteristic time was greater for nanometric obstacles, enhancing in our opinion the possibility to detect dynamics differences when changing only the obstacle orientation. The tilt angle $a$ between the obstacle major axis and the flow has been set to three different values of $47+/-4°$, $63+/-4°$ and $76+/-6°$. Although the hooking probability seems to be highest at $0°$ [13], we just considered these 3 angles due to fabrication constraints linked to our low-cost nanofabrication process [17]. As a reference, our results obtained with cylindrical nanoposts of 270 and 500 nm in diameter were also considered [14]. The minimal distance separating obstacles was ~4.5 µm, so as to minimize entanglement of the DNA with multiple posts during collisions. Solid state matrices were eventually processed as fluidic devices by drilling two inlet holes through silicon, and the resulting chips were sealed on microscope coverslips coated with 5 µm of PDMS after oxygen plasma surface activation [17].

The flow field in the network of obstacles was simulated by finite element modeling using COMSOL in 3D or in 2D with the Hele-Shaw approximation. Both simulations yield similar results, and the latter was shown in Fig 1B. The shallow thickness of our devices of ~0.8 µm, which screens out hydrodynamic interactions, and the small dimension of obstacles of ~0.5 µm in comparison to the distance between consecutive posts of ~4.5 µm, enables to consider obstacles as isolated. The variations of the flow around the obstacle occur over lateral distances of ~500 nm for every configuration (Fig. 1B). Our simulations suggest that the flow velocity $V$ away from the posts is similar for the three tilt configurations. Therefore the Péclet number ($Pe$), which is equal to $Vl_p/D$ with $l_p$ DNA persistence length and $D$ its diffusion coefficient, and which characterizes the deformation of DNA during hooking events [15], is roughly constant for the three lattices at a given migration velocity. The diffusion coefficient of 35 and 49 kbp molecules is 0.5 and 0.4 µm²/s, respectively [14]. Notably the lateral variations of the field around posts occur over much longer distances with hydrodynamic actuation than with electrophoresis, which is associated to sharp field gradients at the edges of the posts (Fig. 1B). Nevertheless, because we are working with molecules of 10 µm or more in this study, we expect these local effects to have minor consequences on unhooking dynamics (see more below).

We then focused on the holdup time of 35 kbp molecules, as defined by the total time of collision events. Using manual tracking of 200 to 600 molecules at a migration velocity of $V$~40 µm/s ($Pe$~4), the distribution of holdup times was extracted for the three elliptical post and compared to a cylindrical obstacle of comparable dimension of 270 nm (Fig. 2A). The tail of histograms is accurately adjusted with an exponential fit, confirming previous results of simulations with electrophoresis [13]. The decay time of the exponential is nearly constant for every type of obstacle on the order of ~200 ms (see results in the inset of Fig. 2A), thus suggesting that the geometry of obstacles has minor effects on DNA hooking dynamics. This effect is associated to the fact that much of the hooked DNA resides far from the obstacle, so the forces triggering its disengagement are marginally affected by the geometry of a small obstacle. Further we investigated long-lived collisions by measuring the unhooking time of at least 10 molecules of 35 or 49 kb at a migration velocity of $V$~20 and 40 µm/s. The unhooking time $t_{unh}$ is defined by the disengagement time after the complete unraveling of the molecule (see inset in Fig. 2B). It can be studied quantitatively using the model of Randall and Doyle [15], which derived a relationship between $t_{unh}$ and DNA conformation at the beginning of the disengagement described by the length of the shorter arm $x_1(0)$ and the end-to-end length $L$ of the molecule (inset in Fig. 2B):

$$t_{unh} = -\frac{\tau_c}{2} ln\left(1 - \frac{2x_1(0)}{L}\right) \qquad (1)$$

with $\tau_c$ the disengagement time, which is $\tau_c = L/V$ for a free draining chain [15]. Figure 2B shows the hooking time as a function of DNA initial conformation, and the dataset are adjusted with equation (1) (bold lines in Fig. 2B, and Supplementary Fig. S1). The disengagement time is given for the three different tilt angles and for a cylindrical post in Table 1. If the dynamics with 63°, 76° and cylindrical obstacles were similar (blue, green, and gray datasets, respectively), an slightly enhanced disengagement time was detected for 47° obstacles. Given the similar profiles of holdup time distributions, we nevertheless conclude that unhooking dynamics are marginally changed by the geometry of nanoposts. Next, we measured the end-to-end lengths of 35 and 49 kbp for each considered event and averaged them in order to calculate the unhooking time $\tau_c$ (table 1). Because the lengths were 16+/-1 or 12+/- 1 µm for 49 and 35 kbp, respectively, the free draining chain model predicts that $\tau_c$ is 0.8+/-0.1 and 0.6+/-0.1 s at 20 µm/s, respectively. The consistency of this model is somewhat surprising because the free draining hypothesis does not apply for the conformation of DNA under hydrodynamic actuation in low-confining geometries (that is, channel height strongly greater than the persistence length of the molecule). In fact, the disengagement time is determined by the balance between viscous *vs.* pulling forces. The viscous drag coefficient is proportional to DNA total length for free-draining chains, but it is dependent on the molecule conformation with hydrodynamic interactions. We recently showed that the shear stress imposed by the Poiseuille flows in confined slit-like channels slows down DNA relaxation, and enhances its deformability during hooking events [14], which adopts a linear elongation, as shown in the inset of Fig. 2B. We thus suggest that each arm can be modeled by a series of homogeneous blobs with minimal inter-blob hydrodynamic interactions, so that the drag coefficient is expected to scale linearly with the molecule length. Consequently the conformational space of hooking events differs in hydrodynamic and electrophoresis [14], but the disengagement kinetics can be described with a free-draining model regardless the actuation principle. Notably the confinement in slit-like geometries appears to be essential for this conclusion, because long-distance hydrodynamic couplings are screened out. This argument is consonant with the results obtained on DNA hooking dynamics in slit channels driven by an electrophoretic force [18], which showed that the viscous drag increased as confinement was enhanced and hence hydrodynamic interactions were screened. As was performed in [18], it would thus be interesting to investigate the difference between hydrodynamics and electrophoresis in the case of channel of ~5 µm in height and with obstacles of different diameters to gain more insights on the difference in formalism between these two actuation principles.

Altogether this study confirms that hydrodynamic actuation provides an efficient solution to manipulate DNA in artificial separation matrices. In addition our single molecule study of DNA disengagement dynamics with elliptical obstacles confirms results of Brownian dynamics simulations, which indicated that DNA unhooking was relatively insensitive to the orientation of the ellipse. We finally ask whether elliptical obstacles are more favorable for DNA separation. For this we evaluated the disengagement time for 35 and 49 kbp molecules as a function of the tilt angle *a*, and observed that the disengagement dynamics was independent on the geometry of the post (Supplementary Fig. S1). Although more data on the behavior of DNA during its migration before and at the onset of the collision is still needed, our work suggests that the challenge associated to fabrication of elliptical nanoposts should not be rewarded by significant improvements in separation performances.

**Acknowledgements** This work was also supported by the LAAS-CNRS technology platform, a member of the French Basic Technology Research Network. Y.V. and H.R. thank the French Defense Procurement Agency and the French Ministry of Research for funding. This work and the fellowship of Q.H. were supported by the ANR program JC08_341867. The authors thank Véronique Langlands for proofreading.

# References

[1] O. Smithies, *Biochemical Journal* 1955, *61*, 629.

[2] B. M. Olivera, P. Baine, N. Davidson, *Biopolymers* 1964, *2*, 245–257.

[3] L. Mitnik, L. Salomé, J. L. Viovy, C. Heller, *Journal of Chromatography A* 1995, *710*, 309–321.

[4] M. N. Albarghouthi, A. E. Barron, *Electrophoresis* 2000, *21*, 4096–4111.

[5] W. D. Volkmuth, R. H. Austin, *Nature* 1992, *358*, 600–602.

[6] O. Bakajin, T. A. J. Duke, J. Tegenfeldt, C.-F. Chou, S. S. Chan, R. H. Austin, E. C. Cox, *Analytical Chemistry* 2001, *73*, 6053–6056.

[7] R. V. Goering, *Infection, Genetics and Evolution* 2010, 10, 866–875.

[8] N. Kaji, Y. Tezuka, Y. Takamura, M. Ueda, T. Nishimoto, H. Nakanishi, Y. Horiike, Y. Baba, *Analytical Chemistry* 2004, *76*, 15–22.

[9] K. D. Dorfman, *Review Modern Physics* 2010, *82*, 2903–2947.

[10] T. Duke, R. Austin, *Physical review letters* 1998, *80*, 1552–1555.

[11] L. R. Huang, E. C. Cox, R. H. Austin, J. C. Sturm, *Analytical chemistry* 2003, *75*, 6963–6967.

[12] Y. C. Chan, Y. Zohar, Y.-K. Lee, *Electrophoresis* 2009, *30*, 3242–3249.

[13] J. Cho, S. Kumar, K. D. Dorfman, *Electrophoresis* 2010, *31*, 860–867.

[14] Y. Viero, Q. He, A. Bancaud, *Small* 2011, *7*, 3508–3518.

[15] G. C. Randall, P. S. Doyle, *Macromolecules* 2006, *39*, 7734–7745.

[16] M. N. Joswiak, J. Ou, K. D. Dorfman, *Electrophoresis* 2012, *33*, 1013–1020.

[17] Y. Viero, Q. He, A. Bancaud, *Microfluidics and Nanofluidics* 2012, *12*, 465–473.

[18] O. B. Bakajin, T. A. J. Duke, C. F. Chou, S. S. Chan, R. H. Austin, E. C. Cox, *Physical Review Letters* 1998, *80*, 2737–2740.

**Figure legends**

**Figure 1:** Fabrication of elliptical nanopost arrays and fluid flow modeling. **(A)** The electron micrographs show arrays of elliptical nanoposts at low magnification (upper panel), or individual nanoposts at high magnification (from left to right, a=47°, 63°, and 76°). The red arrow indicates the orientation of the hydrodynamic field. **(B)** The hydrodynamic flow and the electric field are obtained by 2D finite element modeling for the three obstacles configurations. The maximum flow speed in the interstices of the matrix was set to 40 µm/s. At the bottom, we show the horizontal component of the hydrodynamic flow (left) and electric field (right) magnitude with the velocity field vectors simulated at the single-obstacle level.

**Figure 2:** Single molecule unhooking dynamics. **(A)** The holdup time distribution of U/J collisions is represented for 35 kb molecules and for the three elliptical geometries represented in Figure 1, and for cylindrical nanoposts of 270 nm in diameter. The tail of the distribution is fitted with a single exponential to measure the characteristic decay time of unhooking events. **(B)** The unhooking time is plotted as a function of DNA initial conformation, as recapitulated by the length of the shorter arm $x_1(0)$ at the onset of disengagement, and the end-to-end length $L$ (inset), for 3 different obstacle orientations of 47°, 63°, and 76° (red, green, and blue datasets, respectively), for 35 and 49 kbp molecules (left and right panel, respectively). The bold lines correspond to the fit of the data with the model of Randall and Doyle (Eq. 1). The gray dataset and the corresponding fit are the results obtained with circular posts of 270 nm in diameter.

**Table 1:** Disengagement time for different geometries of obstacles and for two DNA sizes.

|  | Cylindrical | a=47° | a=63° | a=76° | Free draining disengagement time | Chain length |
|---|---|---|---|---|---|---|
| $\tau_c$ (35 kb; Pe~2) | 0.7 +/- 0.1 s | 0.9 +/- 0.2 s | 0.7 +/- 0.1 | 0.6 +/- 0.1 s | 0.6+/-0.1 s | 12+/-1 µm |
| $\tau_c$ (49 kb; Pe~2.4) | 1.1 +/- 0.1 s | 1.2 +/- 0.1 s | 0.9 +/- 0.1 s | 0.9 +/- 0.1 s | 0.8+/-0.1 s | 16+/-1 µm |

**Supplementary Fig. S1:** Ratio of the disengagement dynamics, as deduced in Fig. 2B, of 35 and 49 kb molecules as a function of the tilt angle of the ellipse. The migration velocity is set to 20 µm/s.

Figure 1 Viero et al.

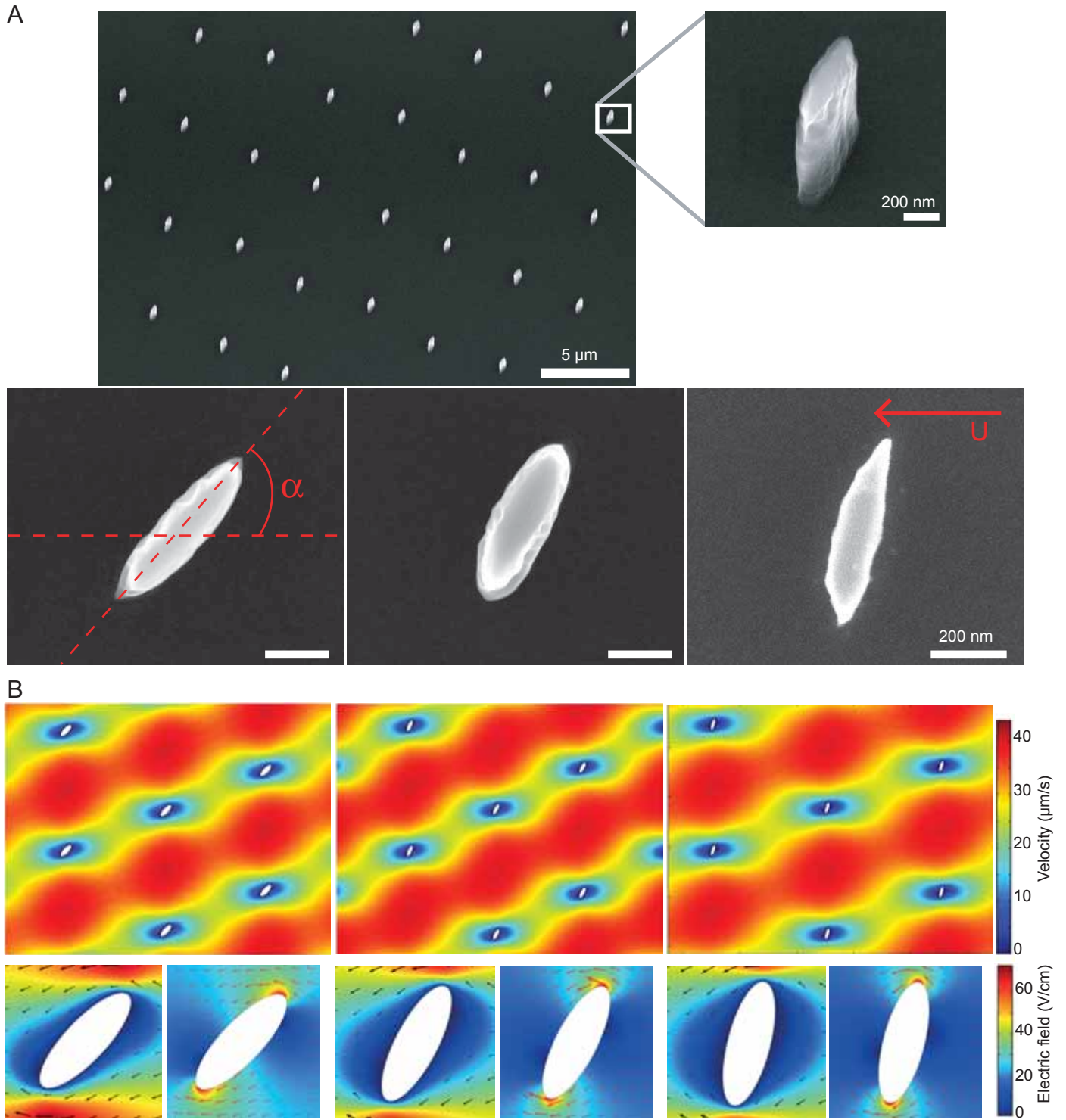

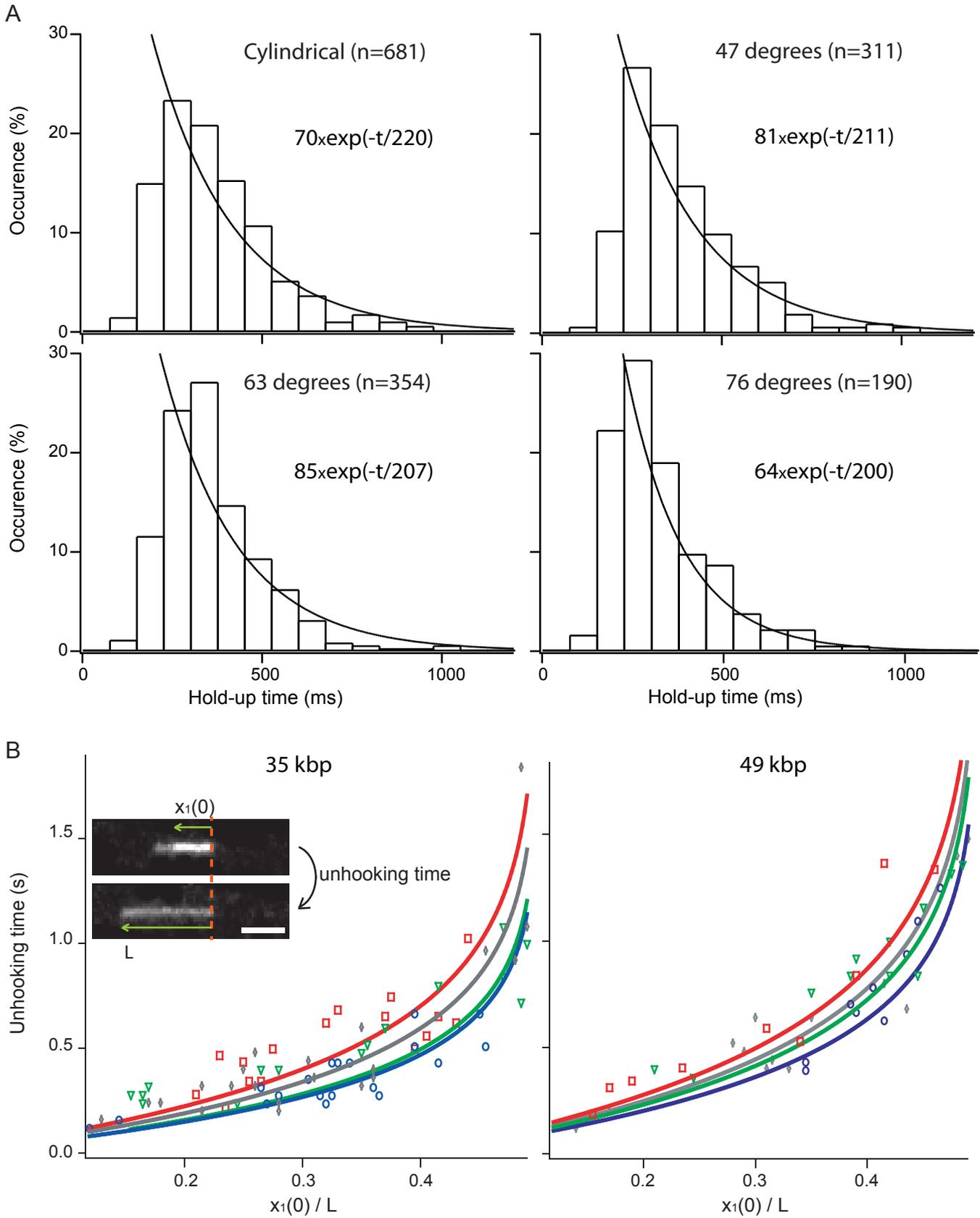